\documentclass[prl,twocolumn,showpacs,amsmath,amssymb]{revtex4}

\usepackage{graphicx}
\usepackage{dcolumn}
\usepackage{bm}

\newcommand{\be}{\begin{equation}}
\newcommand{\ee}{\end{equation}}
\newcommand{\bea}{\begin{eqnarray}}
\newcommand{\eea}{\end{eqnarray}}
\newcommand{\mbf}{\mathbf}
\newcommand{\bs}{\boldsymbol}

\begin{document}

\title{
Quasiparticle density of states of $d$-wave superconductors in a disordered
vortex lattice
}
\author{J. Lages$^1$}
\author{P. D. Sacramento$^1$}
\author{Z. Te\v{s}anovi\'c$^2$}
\affiliation{
$^1$Centro de\! F\'\i sica das\! Interac\c c\~oes Fundamentais,
Instituto Superior T\'ecnico,
Av. Rovisco Pais, 1049-001 Lisboa, Portugal\\
$^2$Department of Physics and Astronomy, Johns Hopkins University, Baltimore,
Maryland 21218, USA
}

\date{\today}

\begin{abstract}
We calculate the density of states of a disordered inhomogeneous $d$-wave
superconductor in a magnetic field. The field-induced vortices are assumed
to be pinned at random positions and the effects of the scattering of
the quasi-particles off the vortices are taken into account using the
singular gauge transformation of Franz and Te\v{s}anovi\'c.
We find two regimes for the density of states: at very low energies the
density of states
follows a law $\rho(\epsilon) \sim \rho_0 + \beta |\epsilon|^{\alpha}$
where the exponent is close to $1$. A good fit of the density of states is obtained
at higher energies, excluding a narrow region around the origin, with a similar
power law energy dependence but with $\alpha$ close to $2$. Both at low and at
higher energies $\rho_0$
scales with the inverse of the magnetic length ($l^{-1} \sim \sqrt{B}$).
\end{abstract}


\maketitle

The effect of disorder on the low-energy
density of states of $d$-wave superconductors
has been a subject of considerable recent interest both for practical and for
theoretical reasons \cite{report}. From the practical
point of view the presence of disorder
pinning mechanisms is important to prevent energy dissipation due to the motion
of the vortices in an external field. From the
theoretical point of view several
conflicting predictions have appeared in the literature.
Some progress toward understanding the
disparity of theoretical results has been achieved
realising that the details of
the type of disorder affect significantly the density of states \cite{report}.
In contrast to conventional gapped $s$-wave
superconductors, the presence of gapless
nodes in $d$-wave superconductors leads to power
law behavior for the low-$T$
thermodynamic properties and is expected to
affect the transport properties.
Linearizing the spectrum around the four Dirac-like nodes it has been suggested
that the system is critical. It was obtained that the density of states is
of the type $\rho(\epsilon) \sim |\epsilon |^{\alpha}$, where $\alpha$
is a non-universal
exponent dependent on the disorder, and that the low
energy modes are extended states
(critical metal) \cite{NTW}. Taking into account the effects of inter-nodal
scattering (hard-scattering) it has been shown that an
insulating state is obtained
instead, where the density of states still vanishes at low energy but with an
exponent $\alpha=1$ independent of disorder \cite{Senthil}.
The addition of time-reversal
breaking creates two new classes designated spin quantum Hall
effect I and II, due
to their similarities to the usual quantum Hall effect,
corresponding to the hard
and soft scattering cases, respectively \cite{report}.
The proposed formation of a
pairing with a symmetry of the type $d+id$ breaks
time-inversion symmetry \cite{did}
but up to now remains a theoretical possibility.
On the other hand applying an external
magnetic field naturally breaks time-reversal invariance and therefore it is
important to study the density of states in this case.

Moreover, the interaction between the superconductor
quasiparticles and the vortices
induced by the external magnetic field has also been a subject of considerable
debate \cite{Gorkov,Anderson,FT}. In the presence
of the vortices the quasiparticles
feel the combined effect of the external magnetic field
and of the spatially varying
field of the chiral supercurrents. Performing a gauge
transformation to effectively
reduce the system to one in a zero average magnetic field
it was shown \cite{FT}
that the natural low energy modes are Bloch waves rather than the Dirac
Landau levels proposed
in \cite{Gorkov,Anderson}. These results also showed that the quasiparticles
besides feeling a Doppler shift caused by the moving supercurrents
\cite{Volovik} (scalar
potential) also feel a quantum ``Berry'' like term due to an
Aharonov-Bohm scattering
of the quasiparticles by the vortices (vector potential).

In this Letter we will be concerned 
with the effect of positional vortex disorder
on the density of states of a 
superconductor in an external magnetic field.
The density of states of a dirty but homogeneous
$s$-wave superconductor in a high
magnetic field where the scattering of 
the quasiparticles off scalar impurities
was considered using a Landau level basis \cite{Dukan}. For small amounts of
disorder it was found that $\rho(\epsilon) \sim \epsilon^2$ but when the
disorder is higher than some critical value a finite density of states is
created
at the Fermi surface. In the same regime of high magnetic fields but with
randomly pinned vortices and no impurities 
the density of states at low energies increases
significantly with respect to the lattice case suggesting a finite value at
zero energy \cite{Sacramento}. 
Refs. \cite{Ye,Khveshchenko} considered the effects of random and
statistically independent scalar and vector potentials
on $d$-wave quasiparticles and it was predicted \cite{Khveshchenko} 
that at low energies $\rho(\epsilon) \sim \rho_0 +a \epsilon^2$,
where $\rho_0 \sim B^{1/2}$.
To our knowledge the effect of randomly pinned {\em discrete}
vortices on the spectrum of a $d$-wave superconductor has not
been addressed previously.

We will consider the lattice
formulation of a disordered $d$-wave superconductor
in a magnetic field. We start from the Bogoliubov-de Gennes (BdG) equations
${\cal H} \psi = \epsilon \psi$
where $\psi^{\dagger}(\mbf{r})=\left(u^*(\mbf{r}),
v^*(\mbf{r}) \right)$ and where the matrix Hamiltonian is given by
\[ {\cal H} = \left( \begin{array}{cc}
\hat{h} & \hat{\Delta} \\
\hat{\Delta}^{\dagger} & -\hat{h}^{\dagger} \\
\end{array} \right)
\]
with \cite{FT,PRB}
\bea
\hat{h} &=& -t \sum_{\bs{\delta}} e^{-\frac{ie}{\hbar c}
\int_{\mbf{r}}^{\mbf{r}+
\mbf{\bs{\delta}}} \mbf{A}(\mbf{r}) \cdot
d\mbf{l}}
\hat{s}_{\bs{\delta}} - \epsilon_F
\nonumber \\
\hat{\Delta} &=& \Delta_0 \sum_{\bs{\delta}} e^{\frac{i}{2}
\phi(\mbf{r})}
\hat{\eta}_{\bs{\delta}} e^{\frac{i}{2} \phi(\mbf{r})}.
\eea
The sums are over nearest neighbors ($\bs{\delta}=\pm
\mbf{x}, \pm \mbf{y}$
on the
square lattice); $\mbf{A}(\mbf{r})$ is the vector
potential,
$\hat{s}_{\bs{\delta}}
u(\mbf{r}) = u(\mbf{r}+\bs{\delta})$,
$\hat{\eta}_{\bs{\delta}}=1/4$ for
$s$-wave
pairing and
$\hat{\eta}_{\bs{\delta}}=(-1)^{\delta_y}
\hat{s}_{\bs{\delta}}$
for $d$-wave pairing. In eq. (1) we have factorized the
phase of the order
parameter and have taken the London limit assuming that the amplitude
$\Delta_0$ is constant
everywhere in space, which is valid in the regime of low fields where the size
of the vortex cores is negligible. It is convenient to perform a singular gauge
transformation to eliminate the phase of the
off-diagonal term in the matrix Hamiltonian.
In such a way the magnetic field is compensated by an array of
opposing half-fluxes. We carry
out the unitary transformation \cite{FT}
${\cal H} \rightarrow {\cal U}^{-1} {\cal H} U$, where
\[ {\cal U} = \left( \begin{array}{cc}
e^{i \phi_A(\mbf{r})} & 0 \\
0 & e^{-i \phi_B(\mbf{r})} \\
\end{array} \right)
\]
with $\phi_A(\mbf{r})+\phi_B(\mbf{r})=\phi
(\mbf{r})$.
The vortices are separated into two groups $A$ and $B$
positioned at $\{\mbf{r}_i^A \}_{i=1,N_A}$ and
$\{\mbf{r}_i^B \}_{i=1,N_B}$
such that the vortices $A$ are only
visible to the particles and the vortices $B$ are only visible to the holes.
The phase fields $\phi_{\mu=A,B}$ are defined through
$\bs{\nabla} \times \bs{\nabla} \phi_{\mu}
(\mbf{r}) = 2 \pi \hat{z}
\sum_i \delta (\mbf{r}-\mbf{r}_i^{\mu})$.
The resulting system is effectively in a magnetic field
$\mbf{B}_{eff}^{\mu} = \mbf{B} - \phi_0 \hat{z}
\sum_i \delta (\mbf{r}-\mbf{r}_i^{\mu})$
that vanishes on average if $N_A=N_B$.
Under the unitary transformation the BdG
equations convert in the $d$-wave case to \cite{PRB}
\begin{widetext}
\bea
-t \sum_{\delta}
e^{i {\cal V}_{\bs{\delta}}^A(\mbf{r})}
u(\mbf{r}+\bs{\delta})
-\epsilon_F u(\mbf{r}) + \Delta_0 \sum_{\delta}
e^{i {\cal A}_{\bs{\delta}}(\mbf{r})}
(-1)^{\delta_y} v(\mbf{r}+\bs{\delta})
&=& \epsilon u(\mbf{r})
\nonumber \\
\Delta_0 \sum_{\delta} e^{i {\cal A}_{\bs{\delta}}
(\mbf{r})}
(-1)^{\delta_y} u(\mbf{r}+\bs{\delta})
+t \sum_{\delta} e^{-i {\cal V}_{\bs{\delta}}^B
(\mbf{r})}
v(\mbf{r}+\bs{\delta})
+\epsilon_F v(\mbf{r})
&=& \epsilon v(\mbf{r})
\eea
\end{widetext}
where the phase factors are given by \cite{PRB}
${\cal V}_{\bs{\delta}}^{\mu}(\mbf{r})
=\int_{\mbf{r}}^{\mbf{r}+\bs{\delta}}
\mbf{k}_s^{\mu}
\cdot d\mbf{l}$ and 
${\cal A}_{\bs{\delta}}(\mbf{r})
=
\frac{1}{2} \int_{\mbf{r}}^{\mbf{r}+
\bs{\delta}}
\left(\mbf{k}_s^A - \mbf{k}_s^B\right)
\cdot d\mbf{l}$. 
Here $\hbar\mbf{k}_s^{\mu} = m \mbf{v}_s^{\mu} =
\hbar\bs{\nabla}
\phi_{\mu} -
\frac{e}{c} \mbf{A}$, where $\mbf{k}_s^{\mu}$
is the superfluid wave vector for the $\mu$-supercurrent.
This quantity can be calculated for an arbitrary configuration
of vortices \cite{PRB} like
\be
\mbf{k}_s^{\mu}(\mbf{r})=2 \pi \int
\frac{d^2k}{(2 \pi)^2}
\frac{i \mbf{k} \times \mbf{z}}{ k^2+\lambda^{-2}}
\sum_{i=1}^{N_\mu} e^{i\mbf{k} \cdot (\mbf{r}-
\mbf{r}_i^{\mu})}.
\ee
Here $\lambda$ is the magnetic penetration length and the sum extends over
all vortex
positions. The BdG equations are then solved taking an
arbitrary configuration of vortices and the density of states
is calculated in the standard way. The average over disorder is then carried
out
calculating the density of states for each vortex configuration and then
performing
the average on the density of states over $100$ different configurations.

\begin{figure}
\includegraphics[width=7cm,height=6.5cm]{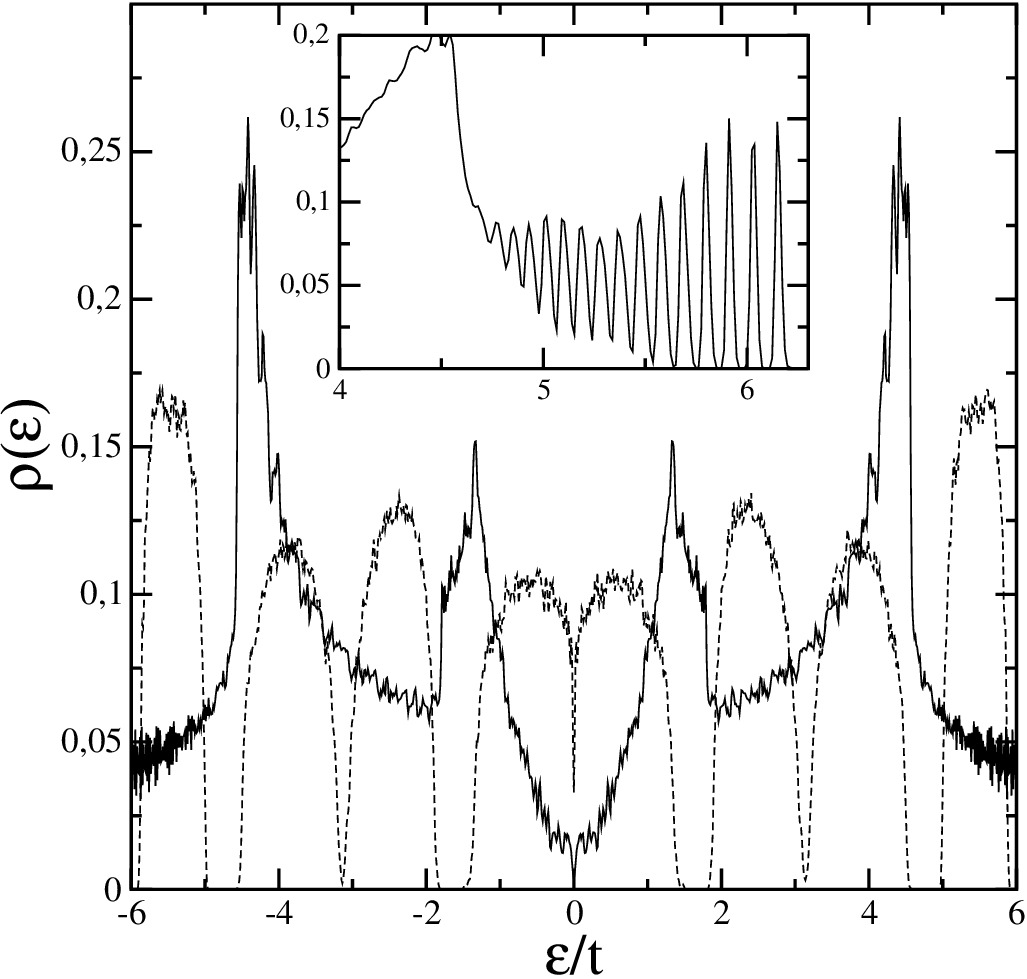}
\caption{\label{fig1} Quasiparticle density of states for different
magnetic fields
$B=1/2$ (dashed line) and $B=1/200$ (solid line) in units of
$hc/(2e\delta^2)$. The inset shows the Landau level quantization
at high-energy for $B=1/50$.
}
\end{figure}

\begin{figure}
\includegraphics[width=7cm,height=6.5cm]{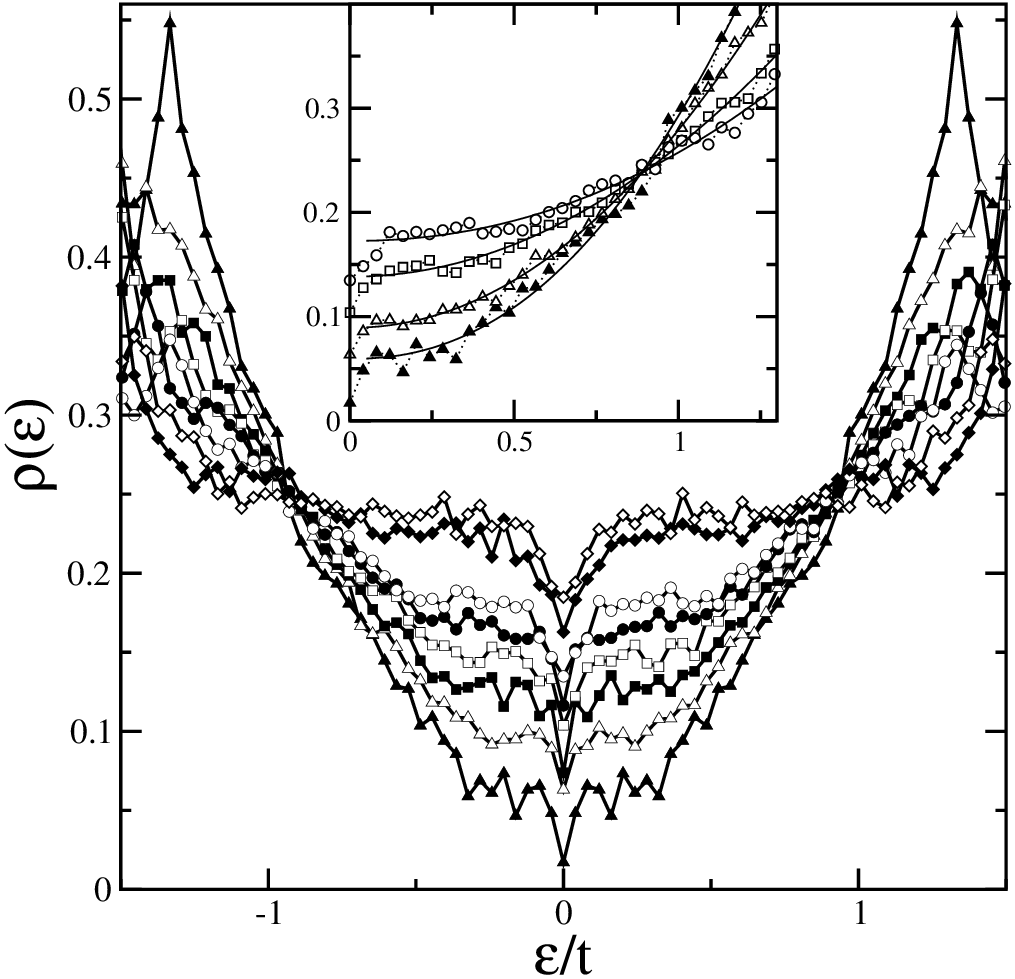}
\caption{\label{fig2}
Quasiparticle density of states for different
magnetic fields
$B=1/200$ ($\blacktriangle$),
$B=3/200$ ($\triangle$),
$B=5/200$ ($\blacksquare$),
$B=7/200$ ($\square$),
$B=9/200$ ($\bullet$),
$B=11/200$ ($\circ$),
$B=20/200$ ($\blacklozenge$)
and
$B=25/200$ ($\lozenge$)
in units of $hc/(2e\delta^2)$.
The linear system size is $L=20$ and the parameters are
$\mu=-2.2t$, $\Delta_0=t$.
The inset shows the fits of the density of states (solid lines)
inside the presented energy interval. For clarity not all the fits are shown.
}
\end{figure}

The situation where the vortices are regularly distributed in a lattice was
treated
before \cite{PRB}. Since the average magnetic field vanishes it is possible
to solve the BdG equations using a standard Bloch basis since the
supercurrent velocities are periodic in space and there is no need to consider
the magnetic Brillouin zone. Taking the continuum limit and linearizing the
spectrum
around each node effectively decouples the nodes.
It was shown that the low-energy
quasiparticles are then naturally described as Bloch waves
and not Dirac-Landau levels as
previously proposed \cite{Gorkov,Anderson}. However, it was found
that in the
linearized problem different
assignments of the $A$ and $B$ vortices lead to
somewhat different spectra, which was
unexpected \cite{PRB}. It was found that taking
the theory on the lattice regularized this problem and
indeed the system has a manifest internal
gauge symmetry such that the spectrum is independent
of the $A$-$B$ vortex assignments, as
it should be. Moreover, the lattice formulation
explicitly involves internodal
contributions which, as discussed above, are
important for the properties of the
density of states in the disordered case.
In the vortex lattice case, however, it was
found that only in special commensurate
cases (for the square lattice) the inclusion of the internodal contributions
is relevant since only in such cases a gap
develops due to the interference
terms between the various nodes, estimated to be of
the order of $\sqrt{B}$. In a general
incommensurate case the interference
is not relevant leading to qualitatively
similar spectra. The spectrum is gapless with a linear
density of states at low energy \cite{PRB}. One would therefore
expect that in a general disordered vortex case
internodal scattering might not be relevant (particularly for
high Dirac cone anisotropy $t/\Delta_0$).

\begin{figure}
\includegraphics[width=7cm,height=6.5cm]{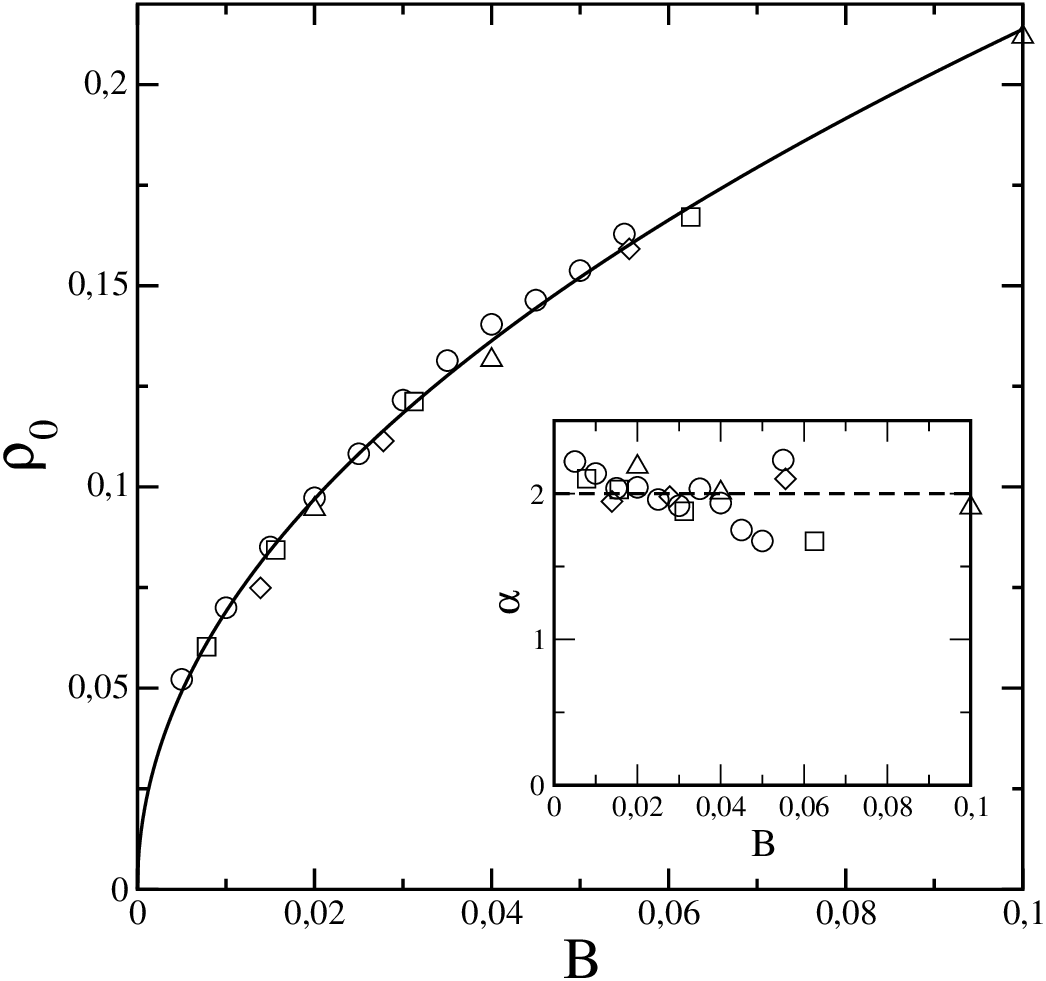}
\caption{\label{fig3} Density of states $\rho_0$ near
the Fermi energy and exponent $\alpha$ (Inset)
extracted from fits for different linear system sizes
$L=10$ ($\triangle$),
$L=12$ ($\diamond$),
$L=16$ ($\square$),
and
$L=20$
($\circ$).
The fit of $\rho_0$ gives a square root dependence in $B$ shown
by the solid line.
}
\end{figure}

We consider a square lattice with lattice constant $\delta$ (taken as $1$),
where the electron hopping is described by
Hamiltonian (1). Application of the external
magnetic field generates vortices which are placed at the
center of a plaquette (unit cell). The number of vortices in the
two-dimensional
system of size $L \times L$ is proportional to the quantized magnetic flux
piercing through the system. We
parametrize the intensity of the magnetic
field by the ratio of the number of
vortices,
$N_{\phi}$, (divided equally in two groups $A$ and $B$)
by the number of lattice sites
$B=\frac{N_{\phi}}{L \times L}$. The $N_{\phi}$ vortices are distributed
randomly over the $L \times L$ plaquettes. We consider then
periodic boundary conditions and solve the BdG equations numerically.
We consider $\lambda \rightarrow \infty$.
In this limit the repulsive interaction between vortices is not screened
and therefore the vortex distribution is not strictly arbitrary. We assume
that the pinning centers are strong enough to overcome the vortex repulsion.
We have checked that the case with a random distribution of vortices
is qualitatively the same as for the case where the vortex
positions are allowed
to vary with a radius of a few unit cells around a
lattice position. We find that, as the disorder
is introduced, the low energy density of states generally increases.

\begin{figure}
\includegraphics[width=7cm,height=6.5cm]{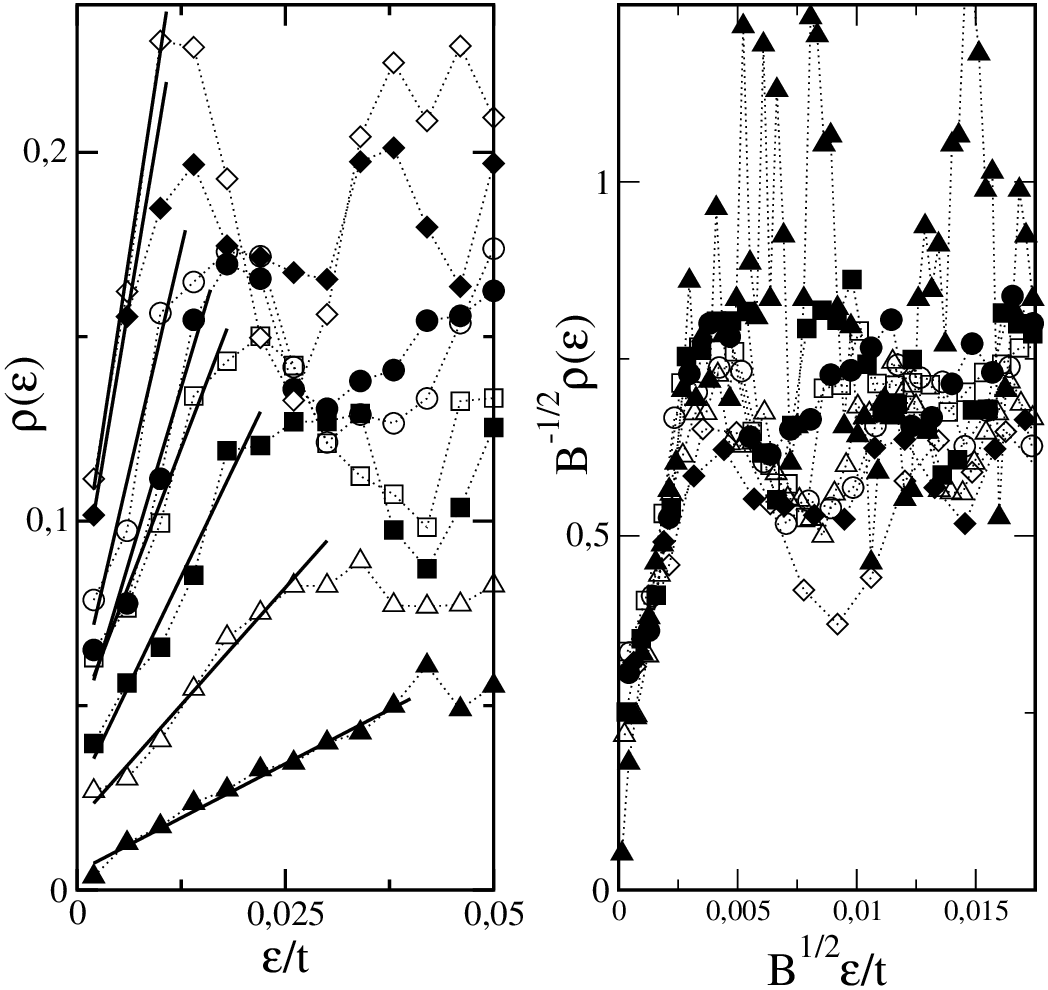}
\caption{\label{fig4} Low-energy quasiparticle density of states
$\rho(\epsilon)$ for different
magnetic field
$B=1/200$ ($\blacktriangle$),
$B=3/200$ ($\triangle$),
$B=5/200$ ($\blacksquare$),
$B=7/200$ ($\square$),
$B=9/200$ ($\bullet$),
$B=11/200$ ($\circ$),
$B=20/200$ ($\blacklozenge$)
and
$B=25/200$ ($\lozenge$)
in unit of $hc/(2e\delta^2)$.
For each data sets the linear system size is $L=20$,
the same as in Fig. \ref{fig2}. In the left panel the solid lines are linear fits
of the dip region below $\epsilon=0.02t$
of the type $\rho (\epsilon) = \rho_{0\rm{dip}}+ \beta | \epsilon |$. In the
right panel we present the near scaling at low energies (see text).
}
\end{figure}

In Figure 1 we show the density of states for the cases of a weak ($B=1/200$)
and of a strong
magnetic field ($B=1/2$; for this high value
the density of vortices is high and
strictly we are in a regime where the size of the vortex cores can not be
neglected). In the case of a strong magnetic field
the Landau level structure at high energies
is clear and it extends to low energies superimposed by the effects of
disorder and the effect of the low-energy modes close to the $d$-wave nodes.
At weak fields the density of states is small at
low energies having a dip close to zero
energy. We have checked for finite size effects on the spectrum. For system
sizes larger than $16 \times 16$ the density of states at not very low energies
converges and the finite size dependence is negligible. 
In Figure 2 we show the density of
states for a system with size $20 \times 20$
and for various magnetic fields. The density
of states at small energies is finite
up to quite small energies where there is a
dip to a value that decreases as the
magnetic field decreases. Only for quite small
magnetic fields the density of states
approaches zero at the origin. Neglecting the
narrow region close to the origin we have
fitted the density of states using the power law
\be
\rho (\epsilon) = \rho_0 + \beta | \epsilon|^{\alpha}.
\ee
In the inset of Figure 2 we show the fits for the various values of the
magnetic field. Reasonable fits are obtained taking $\alpha \sim 2$ and
we obtain that $\rho_0 \sim B^{1/2}$. In Figure 3 we show the magnetic field
dependence of the parameters of the fit for different system sizes
$L=10,12,16,20$.
The various system sizes fit in the same universal curve indicating that the
finite size effects are negligible.
Note that in the lattice case the density of states at low energies
is linear \cite{FT,Marinelli,PRB,AV} (this result differs from
the behavior obtained by others for a $d$-wave superconductor with no
disorder \cite{Volovik,Wang},
where $\rho(\epsilon \sim 0) \sim B^{1/2}$).
The finite density of states at zero energy is therefore a consequence of finite disorder.

\begin{figure}
\includegraphics[width=7cm,height=6.5cm]{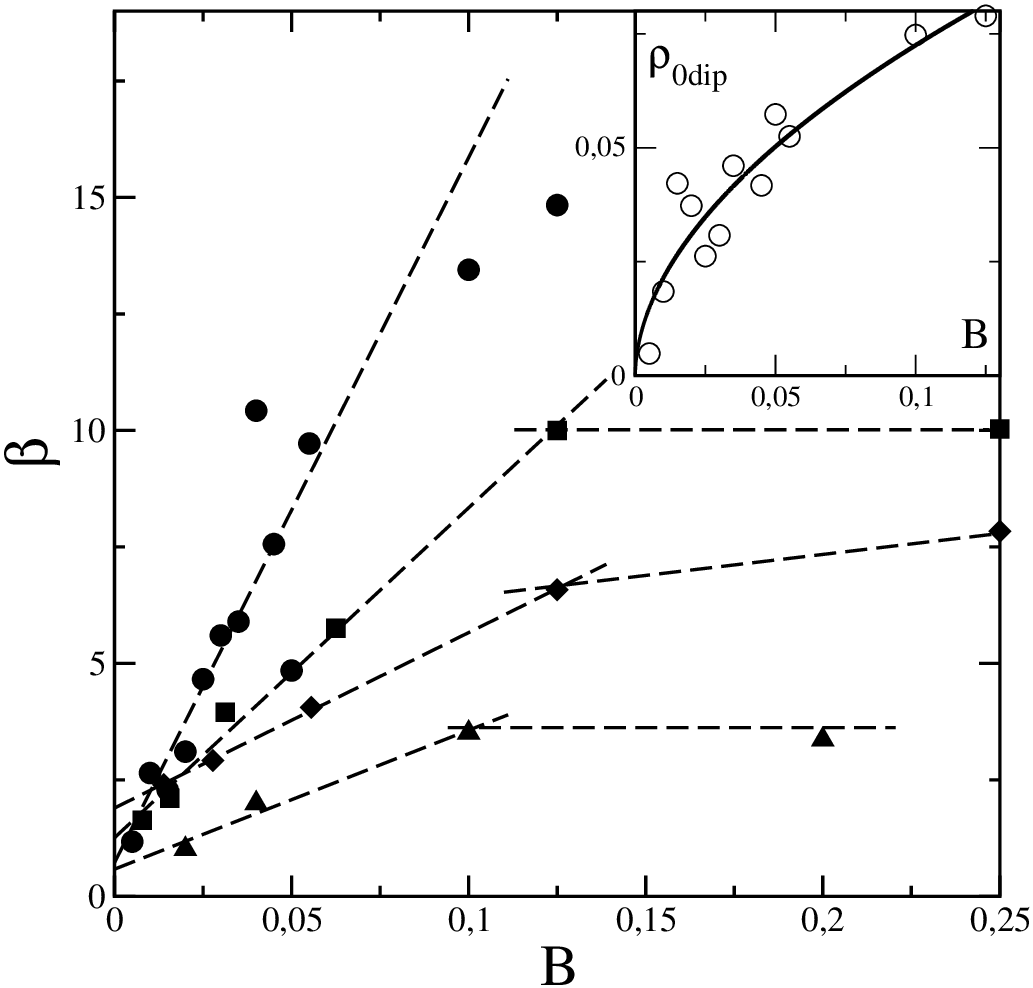}
\caption{\label{fig5}
Slope $\beta$ and density of states $\rho_{0\rm{dip}}$ (Inset)
close to $\epsilon\simeq 0$ extracted in the dip region from data
with different linear lattice sizes $L=10$ ($\blacktriangle$),
$L=12$ ($\blacklozenge$),
$L=16$ ($\blacksquare$)
and $L=20$ ($\bullet$).
Dashed lines are linear fits of $\beta$.
In the inset $\rho_{0\rm{dip}}$ is shown
for $L=20$ ($\circ$).
The fit of $\rho_{0\rm{dip}}$ gives a square root dependence in $B$ shown
by the solid line.
}
\end{figure}

In Figure 4 we focus on the narrow region close to $\epsilon=0$ for the
same set of parameters considered in Figure 2.
Except for the lowest field case $B=1/200$ the
density of states seems to be finite at zero energy. The particular
field density of $B=1/200$ corresponds to only two vortices (since
the size of the system is $20 \times 20$). As shown in \cite{Marinelli}
in this case the spectrum is usually gapped and therefore the density of
states vanishes at zero energy. Performing a fit like in eq. (4) we obtain an
exponent which is now close to $1$.
In this regime $\rho_{0\rm{dip}}$
also scales with $\sqrt{B}$ and the slope scales linearly with $B$.
In this low energy regime the finite size effects are still noticeable
but the dependence on the magnetic field is common to the various
system sizes.
At these low energies the density of states for the various
system sizes appears to be of the following approximate form
$\rho(\epsilon) \sim (1/\omega_H)(1/l^2) {\cal F} \left( (\epsilon/\omega_H)
(\delta^2/l^2) \right)$, where $\omega_H \sim \sqrt{\Delta_0 B}$ and ${\cal F}$ is
a universal function. In the left panel of Figure 4 we show $\rho(\epsilon)$
for various fields while in the right panel
we illustrate the near scaling at low 
energies consistent with ${\cal F}(x) \sim c_1 + c_2 x$ at small $x$.
In Figure 5 we show the field dependence of the slope of $\rho(\epsilon)$
and of the zero energy density of states. Note that $\beta (B\to 0)$ (Fig. 5)
appears to be small but finite, consistent with a crossover to a
``Dirac node'' scaling 
($\rho(\epsilon) \sim (1/\omega_H)(1/l^2) {\cal F} \left(\epsilon/\omega_H\right)$) at very low fields.

In summary, we have calculated the density of states of a disordered $d$-wave
superconductor in a pinned vortex array. Both the disorder
and the magnetic field
fill the density of states at low energies. In general we
find a finite density
of states at zero energy except in the limit of very small magnetic fields.
The density of states deviates from the zero energy value by a power law.
Excluding a narrow region close to zero energy a good fit is obtained with an exponent
close to $2$. Performing the same type of fit at very low energies
a good fit is obtained with an exponent close to $1$.
In the very low magnetic field limit the dip of the density of states is more
pronounced.
The zero energy density of states scales with the inverse of the magnetic
length ($\sqrt{B}$). Except for the zero energy finite value the energy
dependence of the density of states in the case 
with disorder is similar to the
lattice case.
This suggests that the vortex disorder does not dramatically affect the
density of states at low energies. A 
preliminary analysis of the inverse participation
ratio indicates that the states are still extended, as in the lattice case.
Our results for the lowest and zero energies are clearly most susceptible
to finite size effects- we will report on a 
detailed study of this region separately.
In the gapped $s$-wave case however
the disorder introduces states in the gap thereby changing
qualitatively the low energy density of states, as 
in the high field limit \cite{Sacramento}.
These results will be presented elsewhere.

We acknowledge correspondence with D. Khveshchenko 
regarding ref. \cite{Khveshchenko}.
This work was supported in part by NSF grant DMR00-94981 (ZT)
and by FCT Fellowship SFRH/BPD/5602/2001 (JL).

\end{document}